\documentclass[twocolumn,showpacs,superscriptaddress,preprintnumbers,amsmath,amssymb]{revtex4-1}
\usepackage{graphicx}
\usepackage{dcolumn}
\usepackage{appendix}
\usepackage{bm}

\usepackage[usenames,dvipsnames]{xcolor}
\usepackage{subfigure}
\usepackage{bbm}
\usepackage{enumerate}
\usepackage[
  bookmarks=true,
  colorlinks,
  linkcolor=black,
  urlcolor=black,
  citecolor=black,
  plainpages=false,
  pdfpagelabels,
  final,
  breaklinks=true
]{hyperref}

\usepackage{titlesec}
\usepackage{array}
\usepackage{dsfont}
\usepackage{physics}

\begin{document}

\title{On the limitations of the semi-classical picture in high harmonic generation}
\author{Philipp Stammer}
\email{philipp.stammer@icfo.eu}
\affiliation{ICFO -- Institut de Ciencies Fotoniques, The Barcelona Institute of Science and Technology, 08860 Castelldefels (Barcelona), Spain}

\date{\today}
\begin{abstract}

The recent progress in the quantum optical formulation of the process of high harmonic generation has reached a point where the successful semi-classical model shows its limitations.
So far the light source which drives the process was considered to be provided by a laser, in agreement with the classical picture. However, quantum optics allows to consider light fields beyond the classical realm, such as bright squeezed vacuum of photon number states. Both field states have vanishing mean electric field amplitudes, but can still lead to the generation of high harmonic radiation for sufficiently high intensities. 
This poses new questions about the range of validity of the semi-classical picture, and allows to extend the domain of questions which could possibly be asked.

\end{abstract}

\date{\today}

\maketitle

\section{Introduction}

The process of high harmonic generation (HHG) is a frequency upconversion process where an intense light field induces a highly non-linear interaction in matter. This leads to the emission of radiation of very high orders of the driving frequency extending from the infrared to the extreme-ultraviolet regime, and is at the heart of generating attosecond pulses of radiation~\cite{krausz2009attosecond, corkum2007attosecond}.
An intuitive understanding of the underlying electron dynamics is given by the celebrated 3-step model which provides a powerful classical picture for the HHG process~\cite{corkum1993plasma} and allows to obtain qualitative insights about the scattered spectrum. Within this model (i) the electron tunnels through the barrier formed by the combined potential of the Coulomb interaction with the core and the laser electric field coupled to the electron dipole moment. Then, (ii) the electron is driven in the continuum by the presence of the strong electric field and (iii) is eventually driven back to the core and can recombine to it's initial ground state emitting all the excess energy in terms of high energy radiation. 
This idea of (classical) trajectories also appears in the semi-classical description of HHG by means of solving the time-dependent Schrödinger equation for the electron coupled to a classical electric field~\cite{lewenstein1994theory}. The 3-step model is recovered when interpreting the expression of the dipole moment expectation value and provides an electron trajectory approach when using saddle point methods under the strong field approximation~\cite{amini2019symphony}.

Essential in these semi-classical models is the interaction of the electron with the classical oscillating electric field $E_{cl}(t)$. The coupling of the classical field to the electrons dipole moment leads to the emerging tunneling barrier for the electron to tunnel-ionize, and subsequently the electric field drives the electron in the continuum eventually leading to recombination by emitting harmonic radiation. 
This picture suggests that coherent laser radiation with large electric field amplitudes is a key requirement for generating high-order harmonic radiation, and even extending the description of HHG to fully quantized models does not seem to change the picture~\cite{gorlach2020quantum, lewenstein2021generation, stammer2023quantum}. In particular, it was shown, that coupling the electron to the quantized field with an initial coherent state $\ket{\alpha}$ for the driving field leads to coherent states in the harmonic field modes. Furthermore, the coherent state amplitudes of the harmonics are given by the Fourier transform of the time-dependent dipole moment expectation value of the electron -- the same quantity examined in semi-classical models~\cite{lewenstein1994theory}. Thus, the classical features such as coherent harmonic radiation emitted from a classical charge current survives the full quantum description, and notably the semi-classical interaction with $E_{cl}(t)$ is recovered in the case of a coherent laser drive~\cite{lewenstein2021generation}. 

However, the picture of an intense classical field driving the process with an emerging tunneling barrier, and the electron driven by the electric field in the continuum is challenged by recent investigations under the full quantum optical framework~\cite{gorlach2022high}. 
In the following we discuss under which conditions this perspective breaks down and how the picture needs to be re-phrased. 
It turns out that the crucial property of the field determining the existence of the (semi)-classical picture is given by its phase~\cite{stammer2023role}. A consequence of this is that optical coherence is not necessary to drive the process of HHG, and models using the electric field amplitude in strong field dynamics reach their limitations.

\section{Breakdown of the semi-classical picture}

In the semi-classical description of the process of HHG~\cite{lewenstein1994theory} the classical field of the driving laser source $E_{cl}(t)$ is traditionally considered to couple to the electron dipole moment with interaction Hamiltonian 
\begin{align}
    H_{cl}(t) = - d E_{cl}(t).
\end{align}
This leads to the emergence of the tunneling barrier for sufficiently large field amplitudes and further drives the electron in the continuum after ionization (eventually leading to recombination along with the emission of high harmonic radiation).
In order to describe this process within a quantum optical framework~\cite{gorlach2020quantum, lewenstein2021generation, stammer2023quantum} the driving laser of the experimental boundary condition is described by an initial coherent state $\ket{\alpha}$, and the coupling of the charge to the quantized field is taken into account via the electric field operator $E_Q(t)$. The full quantum interaction Hamiltonian acting on the total Hilbert space of the electron and field reads
\begin{align}
    H_Q(t) = - d E_Q(t).
\end{align}

\begin{figure}
    \centering
	\includegraphics[width=1\columnwidth]{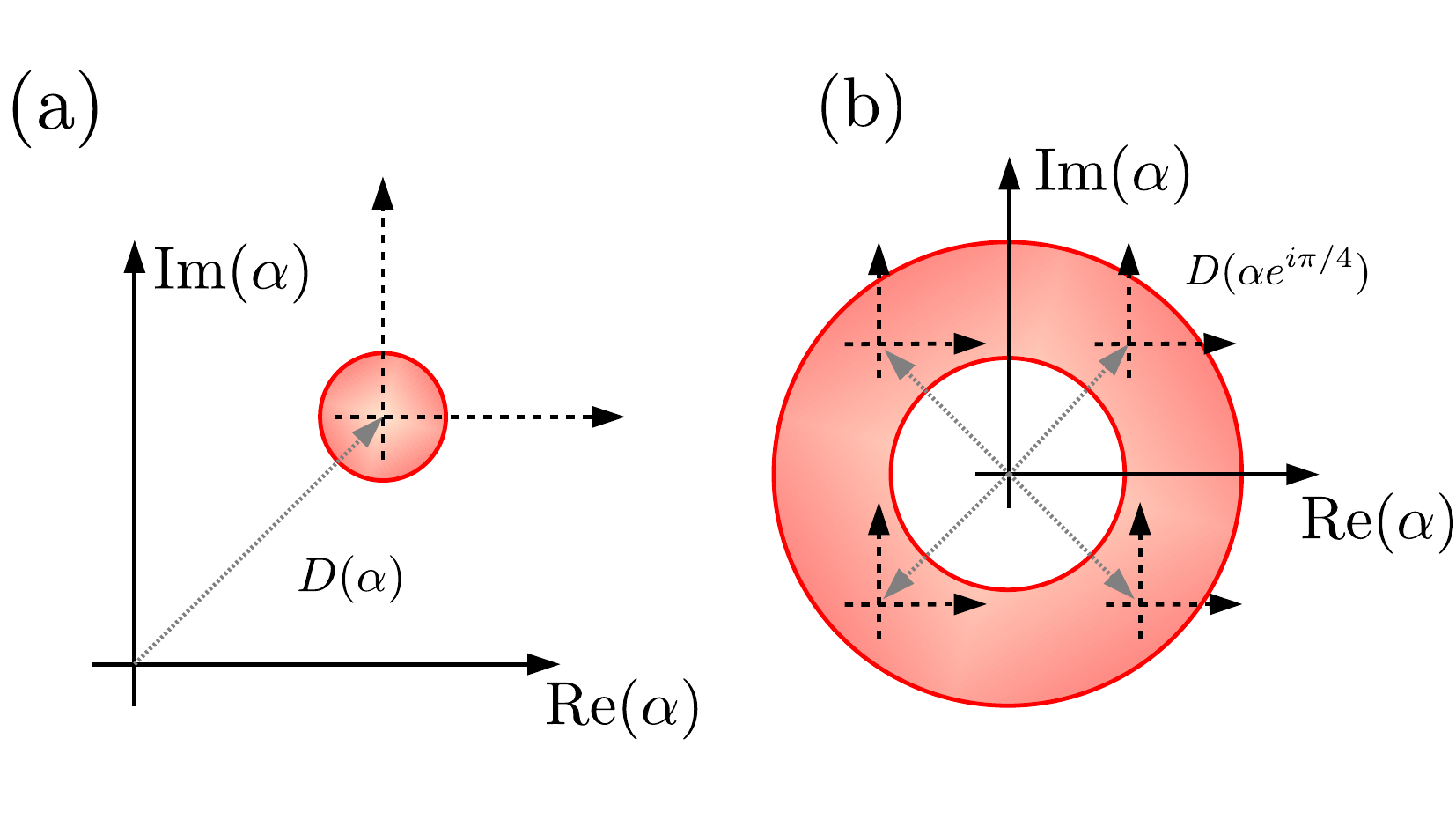}
	\caption{Illustration of the quantum optical phase-space and the emergence of the semi-classical picture via a displacement operation of the reference frame. In (a) for a coherent state driving field $\ket{\alpha}$ such that there exist a unique displacement $D(\alpha)$ recovering the semi-classical picture, while in (b) no unambiguous displacement is possible for the state $\rho_{\abs{\alpha}}$, here exemplified for different displacements such as $D(\alpha e^{i \pi/4})$. In red pictorially the respective states in their phase-space representation.}
     \label{fig:phase_space}
\end{figure}

Taking into account the initial state of the driving laser being in the coherent state $\ket{\alpha }$, a unitary transformation can recover the semi-classical picture. This is done by means of shifting the phase-space reference frame of the initial field state to the vacuum by applying the displacement operator $D(\alpha)$, as illustrated in Fig.~\ref{fig:phase_space}(a).
This leads to the transformed Hamiltonian 
\begin{align}
\label{eq:hamiltonian_total}
    H_I (t) = - d \left[ E_{cl}(t)  +  E_Q(t) \right],
\end{align}
where the semi-classical interaction is recovered, and the new initial state of the driving field is now the vacuum $D^\dagger (\alpha) \ket{\alpha } = \ket{0}$. 
From this Hamiltonian the state of the harmonic field modes $q$ can be obtained and are likewise given by coherent states $\ket{\chi_q}$~\cite{lewenstein2021generation, stammer2023quantum} when dipole moment correlations can be neglected~\cite{stammer2022high, stammer2022theory, stammer2023entanglement}. The harmonic amplitudes $\chi_q$ are proportional to the Fourier transform of the dipole moment expectation value $\expval{d(t)}$, emphasizing the close analogy between the full quantum and semi-classical picture. 
The important aspect to recover the semi-classical interaction in \eqref{eq:hamiltonian_total} is the fact that we can perform a unique unitary transformation with respect to the initial coherent state (see Fig.~\ref{fig:phase_space}). Since the coherent state $\ket{\alpha}$ has a well defined phase $\phi = \operatorname{arg}(\alpha)$, the semi-classical frame via $D(\alpha)$ is unambiguously defined. 
The existence of this well defined semi-classical frame reproduces the classical picture by means of an electron driven by the classical field $E_{cl}(t) = \expval{E_Q(t)} = \Tr[ E_Q(t) \dyad{\alpha }]$. 
This perspective is obtained for the initial coherent state boundary condition describing the state of a classical laser light source.

However, the situation drastically changes when considering different initial conditions for the driving field. In particular, when the process is driven by non-classical light fields, as shown in the recent work by Gorlach et al. \cite{gorlach2022high}.
It was shown that high harmonic radiation can be generated when driving atoms by highly non-classical states of light with notable consequences for the physical picture underlying the semi-classical model. 
The crucial difference in the description of these light states compared to the traditional laser is the absence of a distinct semi-classical reference frame which is only uniquely defined for a coherent state $\ket{\alpha}$ with a well defined phase.

To understand the absence of the unique semi-classical frame, we shall first consider the field of a phase-unstable laser source such that the corresponding state is given by the mixture of coherent states over all possible phases
\begin{align}
\label{eq:state_mixed}
    \rho_{\abs{\alpha}} = \frac{1}{2 \pi} \int_0^{2\pi} d\phi \dyad{\abs{\alpha} e^{i \phi}}.
\end{align}

Since the phase $\phi$ is not well defined for this mixture, there does not exist a unique semi-classical frame via a single unitary displacement operation $D(\abs{\alpha} e^{i \phi})$, as shown in Fig.~\ref{fig:phase_space} (b). Even more important from a physical perspective is the vanishing electric field amplitude for this light state \textit{at all times} 
\begin{align}
    E_{cl} (t) = &  \Tr[E_Q(t) \rho_{\abs{\alpha}}] = 0,
\end{align}
which arise due to the absence of a well defined phase in the field. 
Now focusing on the two non-classical states of light considered in \cite{gorlach2022high}, such as bright squeezed vacuum (BSV) $\ket{\xi}$ or photon number states $\ket{n}$, the same property of a vanishing mean electric field amplitude can be observed for both fields
\begin{align}
    E_{cl}(t) = \Tr[E_Q(t) \dyad{n}] = \Tr[E_Q(t) \dyad{\xi}] = 0.
\end{align}

It is now of interest to see if high harmonic radiation can be generated from these light fields with sufficiently large intensities, and due to the absence of the average field for such light sources, it is natural to ask: \textit{is a non-vanishing electric field amplitude in the driving light necessary for generating high-order harmonic radiation?} 

And indeed, it was shown that driving high-order harmonic generation from sufficiently intense photon number states and bright squeezed vacuum is possible~\cite{gorlach2022high}.
This is because for those field states a vanishing field amplitude does not imply a vanishing field intensity. It is this particular property that the process of HHG can still be driven by those fields, even in the absence of a mean electric field value.  
The interesting insight from this observation is that electric field amplitudes in the driving light are not necessary for the generation of high harmonic radiation. This suggests that the important quantity in strong field dynamics is the field intensity, and not the field amplitude. 
From the classical perspective, the intensity of the field is proportional to the square of the field amplitude $I_{cl} \propto E_{cl}^2$, which would also lead to a vanishing intensity when inconsiderately using $E_{cl}^2 = \expval{E_Q}^2 =0$. 
However, the subtle difference in the quantum optical perspective is that the intensity is obtained from the square of the field operator, and not the square of the expectation value, which for the states under consideration leads to a non-vanishing intensity $I \propto \expval{E_Q^2} \neq 0$.
The disagreement between the two pictures becomes clear when noting that
\begin{align}
\label{eq:intensity_neq_field}
    \expval{E_Q}^2 \neq \expval{E_Q^2}.
\end{align}

Not only does this imply a change in perspective of the underlying picture by means of the 3-step model, but also the respective quantities need to be considered carefully. For instance, in the semi-classical picture the electron has an average kinetic energy due to the motion induced by the oscillating electric field given by the ponderomotive energy $U_p = E_{cl}^2/(4 \omega^2) = \expval{E_Q}^2/(4 \omega^2)$.
But, as we have just seen this would lead to a vanishing quiver motion of the electron in the continuum for light fields with $E_{cl}=0$. 
The subtlety is that despite the vanishing average electric field amplitude, such light fields still have an intensity. It is the intensity of the light which drives the electron leading to the generation of high harmonics. 
The reason why this crucial insight has not received any attention thus far is because for classical light fields described by coherent states with large field amplitudes we have that $\expval{E_Q}^2 \approx \expval{E_Q^2}$, such that the classical perspective is recovered.
Thus, in order to be consistent with the classical and quantum optical perspective, the ponderomotive energy needs to be defined via the average field intensity such that $U_p \propto \expval{E_Q^2}$. This preserves the consistency in the picture for which the ponderomotive energy is non-zero even for light fields with a vanishing electric field, and shows the intrinsic limitations of the semi-classical description which are solely based on using the mean electric field amplitude.

\section{Conclusion}

The growing interest in the quantum optical description of strong laser field induced dynamics has shown advances in several directions from quantum state engineering of light \cite{lewenstein2021generation, stammer2023quantum} to the observation of HHG driven by non-classical light fields \cite{gorlach2022high}. The latter has challenged the understanding of the underlying electron dynamics since the driving light does not require electric field amplitudes for the HHG process.
Despite the power of the semi-classical model to provide an intuitive picture of the underlying process, it poses new questions on the range of it's validity.
It appears that the semi-classical picture has it's limitations not only when driving the process by non-classical light fields, but also for classical states with vanishing mean electric field amplitudes. 
The questions which can now be posed within the quantum optical framework will open up exciting new avenues for strong field driven processes, unforeseen from the semi-classical perspective. Using the field fluctuations of different light sources as an additional degree of freedom to tune the HHG process can lead to an extended cutoff in the spectrum \cite{gorlach2022high} or allows to shape electron trajectories in the continuum \cite{even2023photon}.

In combination with the recent approaches for quantum state engineering of light using the process of high harmonic generation~\cite{stammer2023quantum, pizzi2023light, lamprou2023nonlinear}, this can have promising applications towards quantum information processing using non-classical states of light \cite{lewenstein2022attosecond, bhattacharya2023strong}. Going beyond the classical perspective by investigating its limitations, ultimately connects the two previously unrelated fields of quantum optics and attosecond science \cite{ko2023quantum}.

\begin{acknowledgments}

P.S. acknowledges support from Maciej Lewenstein and the feedback from Nina Meinzer. This work received funding from the European Union’s Horizon 2020 research and innovation programme under the Marie Skłodowska-Curie grant agreement No 847517. 
ICFO group acknowledges support from: Ministerio de Ciencia y Innovation Agencia Estatal de Investigaciones (R$\&$D project CEX2019-000910-S, AEI/10.13039/501100011033, Plan National FIDEUA PID2019-106901GB-I00, FPI), Fundació Privada Cellex, Fundació Mir-Puig, and from Generalitat de Catalunya (AGAUR Grant No. 2017 SGR 1341, CERCA program), and MICIIN with funding from European Union NextGenerationEU(PRTR-C17.I1) and by Generalitat de Catalunya and EU Horizon 2020 FET-OPEN OPTOlogic (Grant No 899794) and ERC AdG NOQIA.
Views and opinions expressed are, however, those of the author(s) only and do not necessarily reflect those of the European Union, European Commission, European Climate, Infrastructure and Environment Executive Agency (CINEA), nor any other granting authority. Neither the European Union nor any granting authority can be held responsible for them.

\end{acknowledgments}

\bibliographystyle{unsrt}
\bibliography{literatur}{}

\end{document}